\documentclass[a4paper]{article}

\usepackage{INTERSPEECH2021}
\usepackage{subfigure}
\title{ConferencingSpeech 2022 Challenge: Non-intrusive Objective Speech Quality Assessment (NISQA) Challenge for Online Conferencing Applications}
\name{Gaoxiong Yi$^1$, Wei Xiao$^1$, Yiming Xiao$^1$, Babak Naderi$^2$, Sebastian Möller$^2$, Wafaa Wardah$^2$, Gabriel Mittag$^3$,Ross Cutler$^3$, Zhuohuang Zhang$^4$, Donald S. Williamson$^4$, Fei Chen$^5$, Fuzheng Yang$^6$, Shidong Shang$^1$}
\address{
  $^1$Tencent Ethereal Audio Lab., China,\\
  $^2$ Technical University of Berlin, Germany,  \quad
   $^3$  Microsoft Corp., USA, \\
    $^4$ Indiana University Bloomington, USA,\\
    $^5$ Southern University of Science and Technology, China,\\
     $^6$ XiDian University, China}
\email{\{gaoxiongyi,denniswxiao,yimingxiao,simeonshang\}@tencent.com$^1$,\{babak.naderi,sebastian.moell
er,wafaa.wardah\}@tu-berlin.de$^2$,\{Ross.Cutler, gmittag\}@microsoft.com$^3$,zhuozhan@iu.edu$^4$, williads@indiana.edu$^4$, fchen@sustech.edu.cn$^5$,fzhyang@mail.xidian.edu.cn$^6$} 
\begin{document}

\maketitle

\section{Abstract}

With the advances in speech communication systems such as online conferencing applications, we can seamlessly work with people regardless of where they are. However, during online meetings, speech quality can be significantly affected by background noise, reverberation, packet loss, network jitter, etc. 
Because of its nature, speech quality is traditionally assessed in subjective tests in laboratories and lately also in crowdsourcing following the international standards from ITU-T Rec. P.800 series. However, those approaches are costly and cannot be applied to customer data. Therefore, an effective objective assessment approach is needed to evaluate or monitor the speech quality of the ongoing conversation. 
The ConferencingSpeech 2022 challenge targets the non-intrusive deep neural network models for the speech quality assessment task. We open-sourced a training corpus with more than 86K speech clips in different languages, with a wide range of synthesized and live degradations and their corresponding subjective quality scores through crowdsourcing. 18 teams submitted their models for evaluation in this challenge. The blind test sets included about 4300 clips from wide ranges of degradations. This paper describes the challenge, the datasets, and the evaluation methods and reports the final results. 

\noindent\textbf{Index Terms}: 
%ConferencingSpeech 2022 challenge, non-intrusive speech quality assessment, mean opinion score
speech quality, deep learning, non-intrusive model

\section{Introduction}

With the popularity of remote conferencing, voice-based human-computer interaction has become mainstream. Environment noise, room reverberation, digital signal processing, and network transmission can all degrade the quality of the speech signal. In these applications, speech quality assessment is in high demand. So far, the above fields have made great progress. 
In ITU-T Rec. P.800 \cite{ong2019}, the international telecommunication union develops subjective evaluation procedures to assess the speech quality, which is the most preferred approach to evaluate the speech quality. However, it must be performed under controlled conditions, which is often time consuming and expensive. Meanwhile, the  perceptual evaluation of speech quality (PESQ) \cite{941023} and perceptual objective listening quality analysis (POLQA) \cite{beerends2013perceptual}  are designed to objectively evaluate speech quality. However, they need clean reference speech signals as comparison input. In order to non-intrusively assess the speech quality,  ITU-T Rec. P.563 \cite{malfait2006p} was developed  but only target narrow-band applications.
As deep learning shines in various fields, deep neural networks have been developed to address
the non-intrusive speech quality assessment problem recently \cite{soni2016novel,spille2018predicting,fu2018quality,andersen2018nonintrusive,lomosnet,gamper2019intrusive,avila2019non,dong2019classification}. Nevertheless, most of these methods adopt PESQ or POLQA as the  speech quality label, which can not really represent the subjective ratings in all impairments. 
Only a few datasets with subjective scores have been published, which limits 
the application of deep learning in the above problem.
Therefore, a large  dataset with subjective speech quality scores and a  non-intrusive speech quality assessment method, which can better reflect perceived subjective feeling, are urgently needed.

The ConferencingSpeech 2022 challenge aims to stimulate research in the above-mentioned areas. We provided comprehensive training and test datasets that contain at least 200 hours of speech samples with subjective test scores.  We hope this challenge helps facilitate idea exchanges and discussions in this special session.  Meanwhile, this challenge has the following features:	1) We aim for non-intrusive models for evaluating the speech quality (i.e., without reference speech signals), which is more practical in online conferencing applications.
2) With the continuous expansion of bandwidth in voice communication systems, the existing standardized non-intrusive objective speech quality assessment method for narrowband speech such as defined in ITU-T P.563 is no longer applicable. Therefore, this challenge aims to effectively evaluate the speech quality for signals with broader bandwidth.
3) To truly reflect subjective opinion on speech quality, the training and test datasets contain the mean opinion score (MOS), which is obtained through  subjective absolute category rating tests in crowdsourcing and in accordance with the ITU-T Rec P.808 \cite{ITU-P808} using its open-sourced implementation \cite{naderi2020open}.
4) As far as we know, this is the first challenge on non-intrusive objective speech quality assessment in online conferencing. We provide a training speech corpus of more than 200 hours of speech samples with corresponding subjective MOS which are covering most of the impairment scenarios users might face in on-line speech communication. It is believed that this will also promote the development of non-intrusive objective speech quality assessment methods.

\begin{table}[tbh]
	\caption{Proportion of the degradations applied in Tencent Corpus.}
	\label{tab:deg:tencent}
	\centering
	\resizebox{\columnwidth}{!}{%
		\begin{tabular}{ l@{}l  c }
			\toprule
			\multicolumn{2}{c}{\textbf{Impairment}} & 
			\multicolumn{1}{c}{\textbf{Percentage}} \\
			\midrule
			White noise                       & & $10\%$~~~             \\
			Nonstationary background noise                      &   & $60\%$~~~               \\
			High-pass/low-pass filtering                       &  & $3.75\%$~~~       \\
			Amplitude clipping                  &   & $1.25\%$~~~              \\
			AMR \cite{bessette2002adaptive}/Opus \cite{valin2012definition} codec                     &   & $5\%$~~~              \\
			Nonstationary background noise + AMR/Opus codec                   &   & $5\%$~~~              \\
			White noise + AMR/Opus codec                &   & $5\%$~~~              \\
			High-pass/low-pass filtering + AMR/Opus codec           & & $5\%$~~~              \\
			Amplitude clipping + non-stationary background noise               &  & $5\%$~~~              \\
			\bottomrule
		\end{tabular}
	}
\end{table}

\section{Task Description}
In this challenge, comprehensive training datasets with ground truth MOS were provided to each registered team. It is anticipated that the participating teams use only the impaired speech signals to design corresponding algorithms or models, so that the output prediction scores are close to the real MOS. The final ranking of this challenge will be determined by the accuracy of the predicted MOS from the submitted model or algorithm on the evaluation test dataset, in terms of root mean squared error (RMSE) and Pearson correlation coefficient (PCC).

It is worth noting that there are no restrictions on the source of the training and development test datasets in this challenge. Participants can use any dataset that is beneficial to the designed algorithm or model for development. However, if additional data is used in training, then an ablation study should be included that shows the benefit to the test set. Meanwhile, the time-consumption and causality of the proposed algorithm or model are not within the scope of this challenge.
\begin{table}[tb]
	\caption{Proportion of the second step simulated impaired speech in Tencent Corpus.}
	\label{tab:final_deg:tencent}
	\centering
	\resizebox{\columnwidth}{!}{%
		\begin{tabular}{ l@{}l  c }
			\toprule
			\multicolumn{2}{c}{\textbf{Impairment}} & 
			\multicolumn{1}{c}{\textbf{Percentage}} \\
			\midrule
			Only first step impairments                    &   & $60\%$ \\
			First step + noise suppression & & $10\%$\\
			First step + noise suppression + packet loss concealment             &   & $1.25\%$~~~              \\
			Clean speech                     &  & $3.75\%$~~~       \\
			Clean speech + packet loss concealment      &   & $1.25\%$~~~              \\
			\bottomrule
		\end{tabular}
	}
\end{table}

\iffalse

\begin{table}[th]
	\caption{Network impairments}
	\label{tab:example}
	\centering
	\begin{tabular}{ c@{}l  c }
		\toprule
		\multicolumn{2}{c}{\textbf{Impairment}} & 
		\multicolumn{1}{c}{\textbf{Percentage}} \\
		\midrule
		Packet loss               & & 40$\%$ $\sim$ 70\%~~~             \\
		jitter              &   & 600 $\sim$ 1200ms~~~               \\
		Throttle (bandwidth limitation)                  &  & 150 $\sim$ 400kb~~~       \\
		\bottomrule
	\end{tabular}
\end{table}

\begin{table}[th]
	\caption{Reverberation parameters}
	\label{tab:example}
	\centering
	\begin{tabular}{ c@{}l  c }
		\toprule
		\multicolumn{2}{c}{\textbf{Room size (m)}} & 
		\multicolumn{1}{c}{\textbf{Reverberation  time (s)}} \\
		\midrule
		$\left[5.4,5.1,2.7\right]$              & & 0.4 ~~~             \\
		$\left[7, 6, 2.7\right]$            &   &0.5 ~~~               \\
		$\left[8, 7, 2.8\right]$              &  & 0.6~~~       \\
		$\left[8, 7, 2.8\right]$                &  & 0.7~~~       \\
		\bottomrule
	\end{tabular}
\end{table}
\fi

\section{Data Description}
In this challenge, we provided the participants with four voice datasets along with MOS labels, namely Tencent Corpus, NISQA Corpus, IU Bloomington Corpus, and PSTN Corpus. Among them, except for the NISQA Corpus, the other three datasets are all made public for the first time. Each dataset will be described in detail below.
\subsection{Tencent Corpus}
This dataset includes speech conditions with reverberation and without reverberation. In the without reverberation condition, there are about 10000 Chinese speech clips and all speech clips experience the simulated impairments which is very often in online conference. In the with reverberation condition, simulated impairments and live recording speech clips are both considered and totally count about 4000. 

In the without reverberation condition, the selected source speech clips were artificially added with some damage to simulate the voice impairment scenario that may be encountered in the online meeting scene. 
In order to prevent the possible speaker-dependent behavior of the trained model, the original speech data was selected from three publicly available datasets  Magic data \cite{Davis80-COP}, ST Mandarin \cite{Rabiner89-ATO} and AIshell\_100h \cite{Hastie09-TEO}. Each speech clip in the source data was processed with one type of impairment and only one type.
The different impairment types and the corresponding percentage of the speech clips applied with each impairment type are listed in Table~\ref{tab:deg:tencent}. Based on the speech clips processed in the above step, we applied another speech processing step including noise suppression \cite{reddy2021icassp} and packet loss concealment \cite{iturecommendation} to simulate more realistic online communication, and also some clean speech clips were added to form the final speech dataset.
Those processing and corresponding percentage in the final dataset are listed in Table~\ref{tab:final_deg:tencent}.

In order to make the subjective database more comprehensive, 4000 speech clips with reverberation were added to the dataset. 28\% of them were generated with simulated reverberation and 72\% were recorded in realistic reverberant rooms. In the simulated reverberation condition, the source data came from the purchased king-asr-166 dataset. Meanwhile, various room sizes and reverberation delays were considered.
The subjective scoring procedure was conducted in a crowdsourcing way similar to ITU-T P.808 including the qualification - training - rating step, except that the training step was simplified due to the scoring platform we were using.
Each clip was rated by more than 24 listeners. After data cleaning, more than 20 subjective scores were obtained for each speech clip and averaged to obtain the final MOS score.

\subsection{NISQA Corpus}
The NISQA Corpus includes more than 14000 speech samples with simulated (e.g., codecs, packet-loss, background noise) and live (e.g., mobile phone, Zoom, Skype, WhatsApp) conditions. The corpus is already publicly available therefore we only included it in the training and development test sets in the competition.
Subjective ratings were collected through an extension of the P.808 Toolkit \cite{naderi2020open} in which participants rated the overall quality and the quality dimensions noisiness, coloration, discontinuity, and loudness. Each clip has on average 5 valid votes. 

Further details about this corpus are provided in \cite{mittag2021nisqa}. 
We also created a new test dataset (TUB hereafter) using unimpaired signals of 136 conversation tests. We selected a portion of speech with no overlapping between two speakers, at least 55\% active speech, and added leading and trailing silences. That leads to 865 source clips, from which a basic clustering algorithm could detect seven different clusters. Meanwhile, we created 62 synthetic degradation conditions (different codecs, bandwidths, single or multiple background noises, packet lost scenarios, etc.). Each condition was applied on seven randomly selected source clips (one per cluster). Finally, we collected on average 18 subjective ratings per clip using the P.808 Framework ~\cite{naderi2020open}.

\subsection{IU Bloomington Corpus}

There are 36000 speech clips extracted from COSINE~\cite{stupakov2009cosine} and VOiCES~\cite{richey2018voices} datasets. The speech clips are truncated between 3 to 6 seconds long, with a total length of around 45 hours.
For the VOiCES dataset, 4 versions of each speech utterance were provided, including reference (i.e., foreground speech), anchor (i.e., low-pass filtered reference), and two reverberant stimuli. The approximated speech-to-reverberation ratios  are between -4.9 to 4.3 dB. Three versions of each speech utterance were provided for the COSINE dataset, including reference (i.e., close-talking mic), anchor, and noisy (i.e., chest or shoulder mic) stimuli. The approximate signal-to-noise ratios (SNRs) range from -10.1 to 11.4 dB.
We crowdsourced our listening tests on Amazon Mechanical Turk by publishing 700 human intelligence tasks  following ITU-R BS.1534 \cite{series2014method}. For more details, please refer to \cite{dong2020pyramid}.

\subsection{PSTN Corpus}
The clean reference files used for the phone calls are derived from the public audiobook dataset Librivox. Because many of the recordings are
of poorer quality, the files have been filtered according to their
quality as described in \cite{reddyinterspeech}, leaving in a total 441 hours from 2150 speakers of good quality speech. These audiobook chapters were then segmented into 10 seconds clips and filtered for having a speech activity of at least $50\%$. Since, in practice, there are often environmental sounds present during phone calls, we used the DNS Challenge 2021 \cite{reddyinterspeech} to add background noise. The noise clips are taken from Audioset \cite{gemmeke2017audio}, Freesound, and the DEMAND \cite{thiemann2013diverse} corpus and added to the clean files with an SNR between $0-40$ dB.
The perceived speech quality of the training and test sets were annotated in a listening experiment on AMT, according to P.808. Each training set file was rated by 5 participants, while the
test set files were rated by 30 participants to ensure a low confidence interval of the MOS values for the model evaluation.  For more details, please refer to \cite{mittag2020dnn}.

\subsection{Dataset Division}
The training, development, and evaluation test sets in this challenge are all originated from the above-mentioned datasets. It is worth noting that the IU Bloomington corpus differs from the Tencent, NISQA and PSTN corpora that used ITU-T P.808 for subjective testing, where the IU Bloomington corpus adopted ITU-R BS.1534 for subjective testing, which resulted in a rating range of 0$\sim$100 instead of 1$\sim$5. Thus, the IU Bloomington corpus will only be provided to participants as additional materials, speech clips from IU Bloomington corpus will not appear in the evaluation test set of the challenge. Participants can decide whether to use it according to their needs. 

Due to the imbalanced size of the datasets, 80\% of Tencent Corpus and 95\% of PSTN Corpus are used for training and development. The rest 20\% of Tencent Corpus, 5\% of PSTN Corpus, and newly created TUB corpus are used for evaluation test in this challenge. We aim to make the impairment situation and score distribution in the divided dataset as even as possible.

In summary, there are about 86000 speech clips for training and development, and 4372 clips for the evaluation test in this challenge. 
They are composed of Chinese, English, and German, and consider background noise, speech enhancement system, reverberation, codecs, packet-loss and other possible online conference voice impairment scenarios.
\iffalse
\section{ Baseline System}
This challenge provided two baseline systems, called Baseline System 1 and Baseline System 2.
The Baseline System 1
is
based on a convolutional neural network (CNN) that estimates the speech quality for each frame of the input signal.
The estimated per-frame quality values are then aggregated
over time by using a recurrent neural network (RNN). The
input to the Baseline System 1 are  transformed to log-mel-spectrograms. To do this, firstly, we calculate spectrograms with
an FFT window length of 1024 samples. We use a hop size
of 480 samples to obtain a time resolution of 10 ms. Then,
a segment of the spectrogram with length of 15 frames (i.e.
150 ms) is extracted, centered around the frame to estimate
the speech quality. After this step, a mel filter bank with a
frequency range from 0 - 16 kHz and 48 bands is applied.
The log energy is then used as input for the CNN. It is worth noting that the Baseline System 1 was not trained on all datasets provided in this challenge, but only on NISQA Corpus。
The overall network structure of the Baseline System 2 is similar to the Baseline System 1. It also inputs the log-mel-spectrograms of the speech signal into the CNN network to extract the speech quality features at different times. But unlike Baseline System 1, attention is used instead of LSTM to fuse speech quality features from different time periods, and training is performed on the entire training set provided in this challenge.
\fi

\section{ Baseline System}
This challenge provided two baseline systems, including Baseline System 1 and Baseline System 2 \cite{mittag2021nisqa}.
The Baseline System 1 is a simplified version of the model in \cite{mittag2019non}. It is made of a deep feed forward network followed by long short-term memory  and average pooling. 
The Baseline System 2 is the complete model from \cite{mittag2021nisqa}. It is based on a convolutional neural network (CNN) and attention mechanism. The log-mel-spectrograms of the speech signal  is provided as input to the CNN network to extract the speech quality features at different times. The estimated per-frame quality values are then aggregated over time by using an attention model and the final score is predicted by the attention pooling block.
Both Baseline Systems were trained on all datasets provided in this challenge.

\section{Challenge Results}
\subsection{Evaluation Setup and Results}
According to ITU-T P.1401~\cite{itu20121401}, we calculated RMSE to evaluate the accuracy, the outlier ratio (OR) for consistency and PCC for linearity. 
%final evaluation was done on the evaluation test set using RMSE and PCC. 

The RMSE is calculated according to the following equation:
\begin{equation}\operatorname{RMSE}=\sqrt{\frac{1}{N-1} \sum_{i=1}^{N} \operatorname{Perror}(i)^{2}},\end{equation}
where $\mathrm{N}$ denotes the total number of speech utterances, $i$ indicates the $i$-th speech signal. Perror represents the prediction error which is defined as the difference between the measured and predicted MOS: 
\begin{equation}
\operatorname{Perror}(i)=\operatorname{MOS}(i)-\operatorname{MOS}_{p}(i).
\end{equation}

The OR represents the number of outlier-points to the total number of speech utterances. An outlier is defined as a point for which the prediction error is larger than the 95\% confidence interval of the MOS value.  
PCC, on the other hand, measures the linear relationship between a
model's estimation and the subjective data.
Meanwhile, due to bias or offsets, different gradients and different qualitative rank orders are always present in subjective evaluations. The statistical uncertainty always exists in the collected MOS. Therefore, a mapping function is recommended to compensate for the possible variance between several subjective experiments. In this challenge, a third-order polynomial function is applied, which can be derived from 
$
y^{\prime \prime}=a+b y+c y^{2}+d y^{3}$, where rmse $\left(x, y^{\prime \prime}\right) \rightarrow \min$ and $f(y)=$ monotonous between $y^{\prime \prime}_{\min}$ and $y^{\prime \prime}_{\max}
$.
\begin{figure}[tb]
	\centering
	
	\includegraphics[width=0.8\columnwidth]{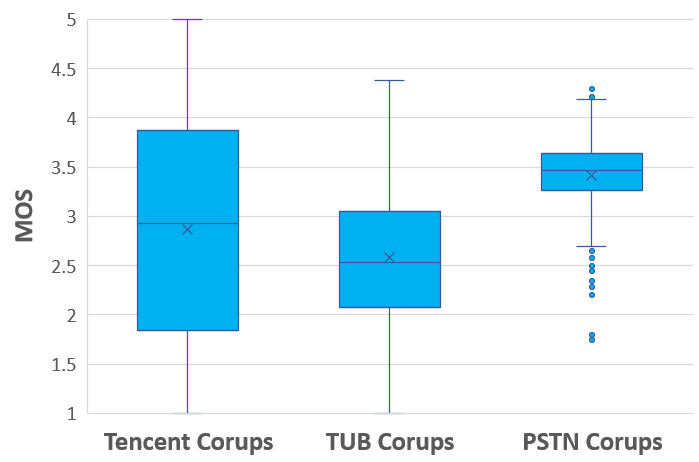}
	%\caption{fig1}
	\caption{ Distribution of MOS values in three blind test sets. }
	\label{fig:blind_set_dist}
\end{figure}

For each model, we create one mapping function per test dataset. Then the mapped predictions were used to calculated RMSE$\_$MAP and OR. In this challenge, we decided to rank models based on their accuracy, i.e., RMSE$\_$MAP. The descriptive statistics on subjective ratings per blind test set are provided in Table~\ref{tab:des:blind} and Figure~\ref{fig:blind_set_dist}.

\begin{table}[tb]
	\caption{Descriptive statistics on subjective ratings in blind test sets.}
	\label{tab:des:blind} 
	\begin{center}
		\resizebox{0.9\columnwidth}{!}{%
			\begin{tabular}{ l c  c  c  c }
				\toprule
				\textbf{Dataset} & \textbf{Average No. } & \textbf{Average}& \multicolumn{2}{c}{\textbf{MOS}}  \\
				& {\textbf{ratings p. clip}}& {\textbf{95\%CI}}&
				{\textbf{min}}&	{\textbf{max}}  \\ 
				\midrule
				Tencent Corpus & 28 & 0.20 & 1.00 & 5.00\\
				TUB Corpus & 18 & 0.40 & 1.00 & 4.37\\
				PSTN Corpus & 24 & 0.35& 1.74 & 4.29 \\
				\bottomrule
			\end{tabular}
		}
	\end{center}
\end{table}

\begin{figure*}[t]
	\centering
	\subfigure[mean of all datasets result analysis]{
		\includegraphics[width=0.35\linewidth]{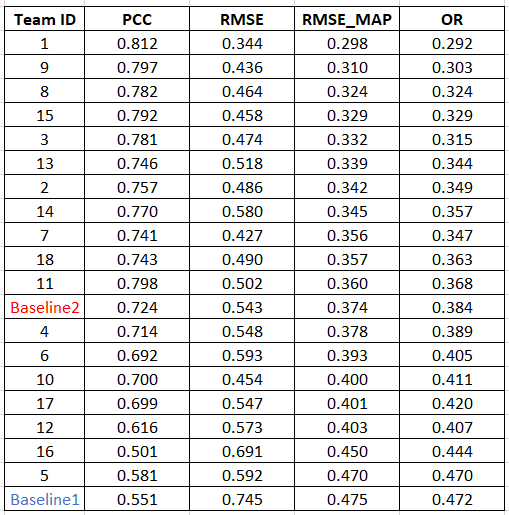}
		%\caption{fig1}
	}\quad
	\subfigure[Tencent Corpus result analysis]{
		\includegraphics[width=0.35\linewidth]{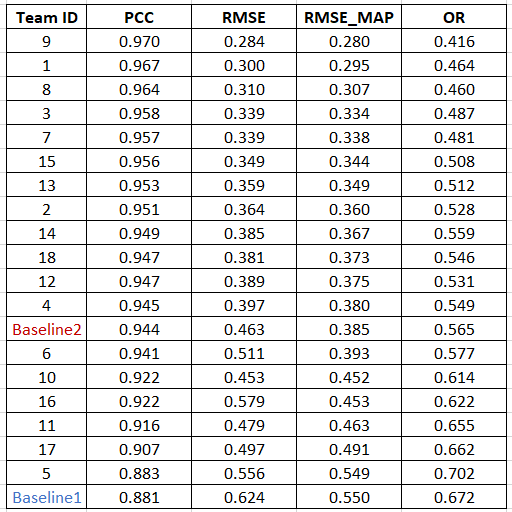}
	}
	\quad
	\subfigure[TUB Corpus result analysis]{
		\includegraphics[width=0.35\linewidth]{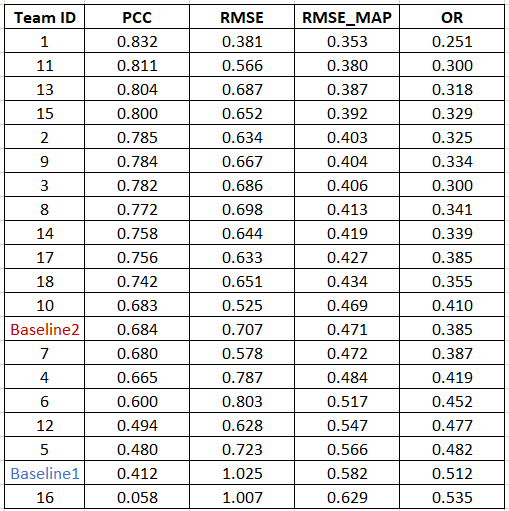}
	}
	\quad
	\subfigure[PSTN Corpus result analysis]{
		\includegraphics[width=0.35\linewidth]{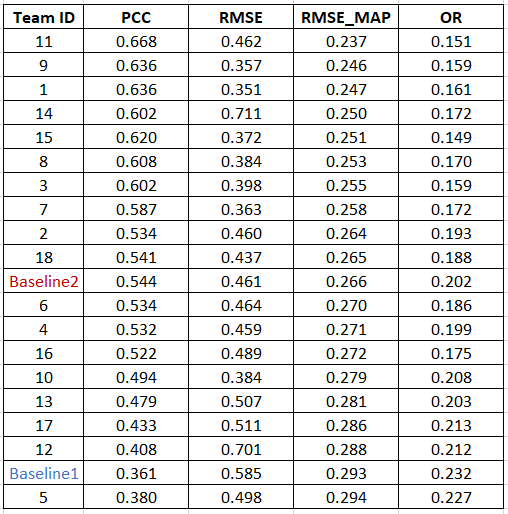}
	}
	\caption{ Challenge result analysis }
	\label{fig:challengeresults}
\end{figure*}

A total of 18 teams from different countries submitted results. Based on the submitted results, their PCC, RMSE, RMSE$\_$MAP, and OR are calculated. The specific results are shown in Figure \ref{fig:challengeresults}. 
\subsection{Key takeaways}
\begin{itemize}
	\item In Figure \ref{fig:challengeresults} (a), it can be observed that the prediction results of all teams are better than the results of the Baseline System 1. It suggests that the generalization of Baseline System 1 is not good enough to effectively cover different datasets. Meanwhile 11 teams achieved better results than the Baseline System 2, accounting for 61$\%$ of the teams who submitted their results.
	%The PCC, RMSE and RMSE$\_$MAP of the best team are 0.800, 0.365 and 0.320, respectively.   
	\item The rank-order of models change based on the dataset and the criteria to use. For any future benchmarking, we recommend to consider multiple blind test sets created by different laboratories. PCC is strongly dependents to how well the MOS values in the test set are distributed: For Tencent Corpus team reach a PCC in range of [0.88 , 0.97] whereas in the other two datasets the achieved PCC was strongly smaller (cf. Figure~\ref{fig:blind_set_dist}). The OR is directly influenced by the 95\% CI, therefore OR values from one dataset cannot be directly compare to another dataset.
	
	%all participating teams had their best prediction results on the Tencent Corpus, with PCC around 90$\%$. The results on TUB are the second, and the PCC can reach from 60$\%$ to 70 $\%$. The worst correlation is found for the PSTN corpus. The PCC can only reach about 40$\%$ to 60$\%$. This may be caused by the number of subjective evaluations of different datasets.
	
	\item We did not observe a significant difference between mapped RMSE values of top 2, 8, and 9 models in datasets Tencent, TUB, and PSTN, respectively. However, team \#1 is consistently in the top-three for all datasets.
	
\end{itemize}
\section{Conclusion}
The ConferencingSpeech 2022 Challenge was organized to help researchers from academia and industry to facilitate the development of non-intrusive objective speech quality assessment for online conferencing applications. We open-sourced several large training datasets with subjective scores. We recommend considering different blind test sets, created by multiple groups, for any similar challenges or benchmarking tasks.
%Many participants from both industry and academia found the datasets very beneficial.

\bibliographystyle{IEEEtran}

\bibliography{mybib}

\begin{thebibliography}{10}
\providecommand{\url}[1]{#1}
\csname url@samestyle\endcsname
\providecommand{\newblock}{\relax}
\providecommand{\bibinfo}[2]{#2}
\providecommand{\BIBentrySTDinterwordspacing}{\spaceskip=0pt\relax}
\providecommand{\BIBentryALTinterwordstretchfactor}{4}
\providecommand{\BIBentryALTinterwordspacing}{\spaceskip=\fontdimen2\font plus
\BIBentryALTinterwordstretchfactor\fontdimen3\font minus
  \fontdimen4\font\relax}
\providecommand{\BIBforeignlanguage}[2]{{%
\expandafter\ifx\csname l@#1\endcsname\relax
\typeout{** WARNING: IEEEtran.bst: No hyphenation pattern has been}%
\typeout{** loaded for the language `#1'. Using the pattern for}%
\typeout{** the default language instead.}%
\else
\language=\csname l@#1\endcsname
\fi
#2}}
\providecommand{\BIBdecl}{\relax}
\BIBdecl

\bibitem{ong2019}
{ITU-T Recommendation P.800}, \emph{{Methods for subjective determination of
  transmission quality}}.\hskip 1em plus 0.5em minus 0.4em\relax Geneva:
  International Telecommunication Union, 1996.

\bibitem{941023}
A.~Rix, J.~Beerends, M.~Hollier, and A.~Hekstra, ``Perceptual evaluation of
  speech quality ({PESQ})-a new method for speech quality assessment of
  telephone networks and codecs,'' in \emph{2001 IEEE International Conference
  on Acoustics, Speech, and Signal Processing.}, vol.~2, 2001, pp. 749--752
  vol.2.

\bibitem{beerends2013perceptual}
J.~G. Beerends, C.~Schmidmer, J.~Berger, M.~Obermann, R.~Ullmann, J.~Pomy, and
  M.~Keyhl, ``Perceptual objective listening quality assessment ({POLQA}), the
  third generation {ITU-T} standard for end-to-end speech quality measurement
  part {I}—temporal alignment,'' \emph{Journal of the Audio Engineering
  Society}, vol.~61, no.~6, pp. 366--384, 2013.

\bibitem{malfait2006p}
L.~Malfait, J.~Berger, and M.~Kastner, ``P.563—the {ITU-T} standard for
  single-ended speech quality assessment,'' \emph{IEEE Transactions on Audio,
  Speech, and Language Processing}, vol.~14, no.~6, pp. 1924--1934, 2006.

\bibitem{soni2016novel}
M.~H. Soni and H.~A. Patil, ``Novel deep autoencoder features for non-intrusive
  speech quality assessment,'' in \emph{2016 24th European Signal Processing
  Conference (EUSIPCO)}.\hskip 1em plus 0.5em minus 0.4em\relax IEEE, 2016, pp.
  2315--2319.

\bibitem{spille2018predicting}
C.~Spille, S.~D. Ewert, B.~Kollmeier, and B.~T. Meyer, ``Predicting speech
  intelligibility with deep neural networks,'' \emph{Computer Speech \&
  Language}, vol.~48, pp. 51--66, 2018.

\bibitem{fu2018quality}
S.~Fu, Y.~Tsao, H.~Hwang, and H.~Wang, ``{Quality-Net:} an end-to-end
  non-intrusive speech quality assessment model based on {BLSTM},'' in
  \emph{Proc. Interspeech 2018}, 2018.

\bibitem{andersen2018nonintrusive}
A.~H. Andersen, J.~M. De~Haan, Z.-H. Tan, and J.~Jensen, ``Nonintrusive speech
  intelligibility prediction using convolutional neural networks,''
  \emph{IEEE/ACM Transactions on Audio, Speech, and Language Processing},
  vol.~26, no.~10, pp. 1925--1939, 2018.

\bibitem{lomosnet}
C.-C. Lo, S.-W. Fu, W.-C. Huang, X.~Wang, J.~Yamagishi, Y.~Tsao, and H.-M.
  Wang, ``{MOSNet}: Deep learning-based objective assessment for voice
  conversion,'' in \emph{Proc. Interspeech 2019}, 2019.

\bibitem{gamper2019intrusive}
H.~Gamper, C.~K. Reddy, R.~Cutler, I.~J. Tashev, and J.~Gehrke, ``Intrusive and
  non-intrusive perceptual speech quality assessment using a convolutional
  neural network,'' in \emph{2019 IEEE Workshop on Applications of Signal
  Processing to Audio and Acoustics (WASPAA)}.\hskip 1em plus 0.5em minus
  0.4em\relax IEEE, 2019, pp. 85--89.

\bibitem{avila2019non}
A.~R. Avila, H.~Gamper, C.~Reddy, R.~Cutler, I.~Tashev, and J.~Gehrke,
  ``Non-intrusive speech quality assessment using neural networks,'' in
  \emph{ICASSP 2019-2019 IEEE International Conference on Acoustics, Speech and
  Signal Processing (ICASSP)}.\hskip 1em plus 0.5em minus 0.4em\relax IEEE,
  2019, pp. 631--635.

\bibitem{dong2019classification}
X.~Dong and D.~S. Williamson, ``A classification-aided framework for
  non-intrusive speech quality assessment,'' in \emph{2019 IEEE Workshop on
  Applications of Signal Processing to Audio and Acoustics (WASPAA)}.\hskip 1em
  plus 0.5em minus 0.4em\relax IEEE, 2019, pp. 100--104.

\bibitem{ITU-P808}
{ITU-T Recommendation P.808}, \emph{{Subjective evaluation of speech quality
  with a crowdsourcing approach}}.\hskip 1em plus 0.5em minus 0.4em\relax
  Geneva: International Telecommunication Union, 2021.

\bibitem{naderi2020open}
B.~Naderi and R.~Cutler, ``{An Open Source Implementation of ITU-T
  Recommendation P.808 with Validation},'' in \emph{Proc. Interspeech 2020},
  2020.

\bibitem{bessette2002adaptive}
B.~Bessette, R.~Salami, R.~Lefebvre, M.~Jelinek, J.~Rotola-Pukkila, J.~Vainio,
  H.~Mikkola, and K.~Jarvinen, ``The adaptive multirate wideband speech codec
  (amr-wb),'' \emph{IEEE transactions on speech and audio processing}, vol.~10,
  no.~8, pp. 620--636, 2002.

\bibitem{valin2012definition}
J.-M. Valin, K.~Vos, and T.~Terriberry, ``Definition of the opus audio codec,''
  \emph{IETF, September}, vol.~2, 2012.

\bibitem{Davis80-COP}
``Magic data,'' \url{https://www.magicdatatech.cn/datasets}, accessed:
  2022-02-25.

\bibitem{Rabiner89-ATO}
``Slr38: Free st chinese mandarin corpus,'' \url{http://www.openslr.org/38/},
  accessed: 2022-02-25.

\bibitem{Hastie09-TEO}
``Aishell- open source mandarin speech corpus,''
  \url{http://www.aishelltech.com/kysjcp}, accessed: 2022-02-25.

\bibitem{reddy2021icassp}
C.~K. Reddy, H.~Dubey, V.~Gopal, R.~Cutler, S.~Braun, H.~Gamper, R.~Aichner,
  and S.~Srinivasan, ``Icassp 2021 deep noise suppression challenge,'' in
  \emph{ICASSP 2021-2021 IEEE International Conference on Acoustics, Speech and
  Signal Processing (ICASSP)}.\hskip 1em plus 0.5em minus 0.4em\relax IEEE,
  2021, pp. 6623--6627.

\bibitem{iturecommendation}
T.~ITU, ``Recommendation g. 191,(2005),'' \emph{Software Tools for Speech and
  Audio Coding Standardization}.

\bibitem{mittag2021nisqa}
G.~Mittag, B.~Naderi, A.~Chehadi, and S.~Möller, ``{NISQA}: A deep
  cnn-self-attention model for multidimensional speech quality prediction with
  crowdsourced datasets,'' in \emph{Proc. Interspeech 2021}, 2021.

\bibitem{stupakov2009cosine}
A.~Stupakov, E.~Hanusa, J.~Bilmes, and D.~Fox, ``Cosine-a corpus of multi-party
  conversational speech in noisy environments,'' in \emph{2009 IEEE
  International Conference on Acoustics, Speech and Signal Processing}.\hskip
  1em plus 0.5em minus 0.4em\relax IEEE, 2009, pp. 4153--4156.

\bibitem{richey2018voices}
C.~Richey, M.~A. Barrios, Z.~Armstrong, C.~Bartels, H.~Franco, M.~Graciarena,
  A.~Lawson, M.~K. Nandwana, A.~Stauffer, J.~{van Hout}, P.~Gamble,
  J.~Hetherly, C.~Stephenson, and K.~Ni, ``Voices obscured in complex
  environmental settings {(VOiCES)} corpus,'' in \emph{Proc. Interspeech 2018},
  2018.

\bibitem{series2014method}
B.~Series, ``Method for the subjective assessment of intermediate quality level
  of audio systems,'' \emph{International Telecommunication Union
  Radiocommunication Assembly}, 2014.

\bibitem{dong2020pyramid}
X.~Dong and D.~S. Williamson, ``{A Pyramid Recurrent Network for Predicting
  Crowdsourced Speech-Quality Ratings of Real-World Signals},'' in \emph{Proc.
  Interspeech 2020}, 2020.

\bibitem{reddyinterspeech}
C.~K. Reddy, V.~Gopal, R.~Cutler, E.~Beyrami, R.~Cheng, H.~Dubey,
  S.~Matusevych, R.~Aichner, A.~Aazami, S.~Braun, P.~Rana, S.~Srinivasan, and
  J.~Gehrke, ``{The INTERSPEECH 2020 Deep Noise Suppression Challenge}:
  Datasets, subjective testing framework, and challenge results,'' in
  \emph{Proc. Interspeech 2020}, 2020.

\bibitem{gemmeke2017audio}
J.~F. Gemmeke, D.~P. Ellis, D.~Freedman, A.~Jansen, W.~Lawrence, R.~C. Moore,
  M.~Plakal, and M.~Ritter, ``Audio set: An ontology and human-labeled dataset
  for audio events,'' in \emph{2017 IEEE International Conference on Acoustics,
  Speech and Signal Processing (ICASSP)}.\hskip 1em plus 0.5em minus
  0.4em\relax IEEE, 2017, pp. 776--780.

\bibitem{thiemann2013diverse}
J.~Thiemann, N.~Ito, and E.~Vincent, ``The diverse environments multi-channel
  acoustic noise database (demand): A database of multichannel environmental
  noise recordings,'' in \emph{Proceedings of Meetings on Acoustics ICA2013},
  vol.~19, no.~1.\hskip 1em plus 0.5em minus 0.4em\relax Acoustical Society of
  America, 2013, p. 035081.

\bibitem{mittag2020dnn}
G.~Mittag, R.~Cutler, Y.~Hosseinkashi, M.~Revow, S.~Srinivasan, N.~Chande, and
  R.~Aichner, ``{DNN} no-reference {PSTN} speech quality prediction,'' in
  \emph{Proc. Interspeech 2020}, 2020.

\bibitem{mittag2019non}
G.~Mittag and S.~M{\"o}ller, ``Non-intrusive speech quality assessment for
  super-wideband speech communication networks,'' in \emph{ICASSP 2019-2019
  IEEE International Conference on Acoustics, Speech and Signal Processing
  (ICASSP)}.\hskip 1em plus 0.5em minus 0.4em\relax IEEE, 2019, pp. 7125--7129.

\bibitem{itu20121401}
{ITU-T Recommendation P.1401}, \emph{{Methods, metrics and procedures for
  statistical evaluation, qualification and comparison of objective quality
  prediction models}}.\hskip 1em plus 0.5em minus 0.4em\relax Geneva:
  International Telecommunication Union, 2020.

\end{thebibliography}
\end{document}